\pdfoutput=1

\documentclass[12pt]{article}
\usepackage{graphicx}
\usepackage{amsmath}
\usepackage{cite}

\newcommand\pubnumber{DPF2013-45}
\newcommand\pubdate{September 30, 2013}

\def\warwick{Department of Physics\\
University of Warwick, Coventry, UK}
\def\support{\footnote{On behalf of the LHCb Collaboration.}}

\def\Title#1{\begin{center} {\Large #1 } \end{center}}
\def\Author#1{\begin{center}{ \sc #1} \end{center}}
\def\Address#1{\begin{center}{ \it #1} \end{center}}

\newcommand\pubblock{\rightline{\begin{tabular}{l} \pubnumber\\
         \pubdate  \end{tabular}}}
\newenvironment{Abstract}{\begin{quotation}  }{\end{quotation}}
\newenvironment{Presented}{\begin{quotation} \begin{center} 
             PRESENTED AT\end{center}\bigskip 
      \begin{center}\begin{large}}{\end{large}\end{center} \end{quotation}}
\def\Acknowledgments{\bigskip  \bigskip \begin{center} \begin{large}
             \bf ACKNOWLEDGMENTS \end{large}\end{center}}

\def\beq{\begin{equation}}
\def\eeq#1{\label{#1}\end{equation}}
\def\eeqn{\end{equation}}

\def\beqa{\begin{eqnarray}}
\def\eeqa#1{\label{#1}\end{eqnarray}}
\def\eeqan{\end{eqnarray}}

\let\bar=\overbar

\def\Dslash{\not{\hbox{\kern-4pt $D$}}}
\def\dslash{\not{\hbox{\kern-2pt $\del$}}}

\def\msb{{\bar{\ssstyle M \kern -1pt S}}}

\begin{document}
\begin{titlepage}
\pubblock

\vfill
\Title{Measurements of $B \rightarrow DK$ decays to constrain the CKM Unitarity Triangle angle $\gamma$ and related results at LHCb}
\vfill
\Author{ Daniel Craik \support}
\Address{\warwick}
\vfill
\begin{Abstract}
Constraints on the CKM angle $\gamma$ are presented from GLW, ADS, and GGSZ analyses of $B^\pm \rightarrow D K^\pm$ at the LHCb experiment. 
The branching fractions of $B^0 \rightarrow \bar{D}^0 K^+ \pi^-$ and $B_s^0 \rightarrow \bar{D}^0 K^- \pi^+$ are also 
reported, measured relative to the related mode $B^0 \rightarrow \bar{D}^0 \pi^+ \pi^-$.
\end{Abstract}
\vfill
\begin{Presented}
DPF 2013\\
The Meeting of the American Physical Society\\
Division of Particles and Fields\\
Santa Cruz, California, August 13--17, 2013\\
\end{Presented}
\vfill
\end{titlepage}
\def\thefootnote{\fnsymbol{footnote}}
\setcounter{footnote}{0}

\section{Measurements of $\gamma$ from $B^\pm \rightarrow D K^\pm$}
\label{sec:gamma}

The CKM angle $\gamma=\text{arg}(-V_{\text{ud}}V^*_{\text{ub}}/V_{\text{cd}}V^*_{\text{cb}})$ is currently the least well-constrained angle in the Unitarity Triangle. 
So far, the most-sensitive measurements of $\gamma$ from a single experiment have been performed by Belle~\cite{Trabelsi:2013uj} and BaBar~\cite{Lees:2013zd}. These measurements yield values of $\left(68^{+15}_{-14}\right)^\circ$ and $\left(69^{+17}_{-16}\right)^\circ$, respectively.

\begin{figure}[htb]
\centering
\includegraphics[height=1.4in]{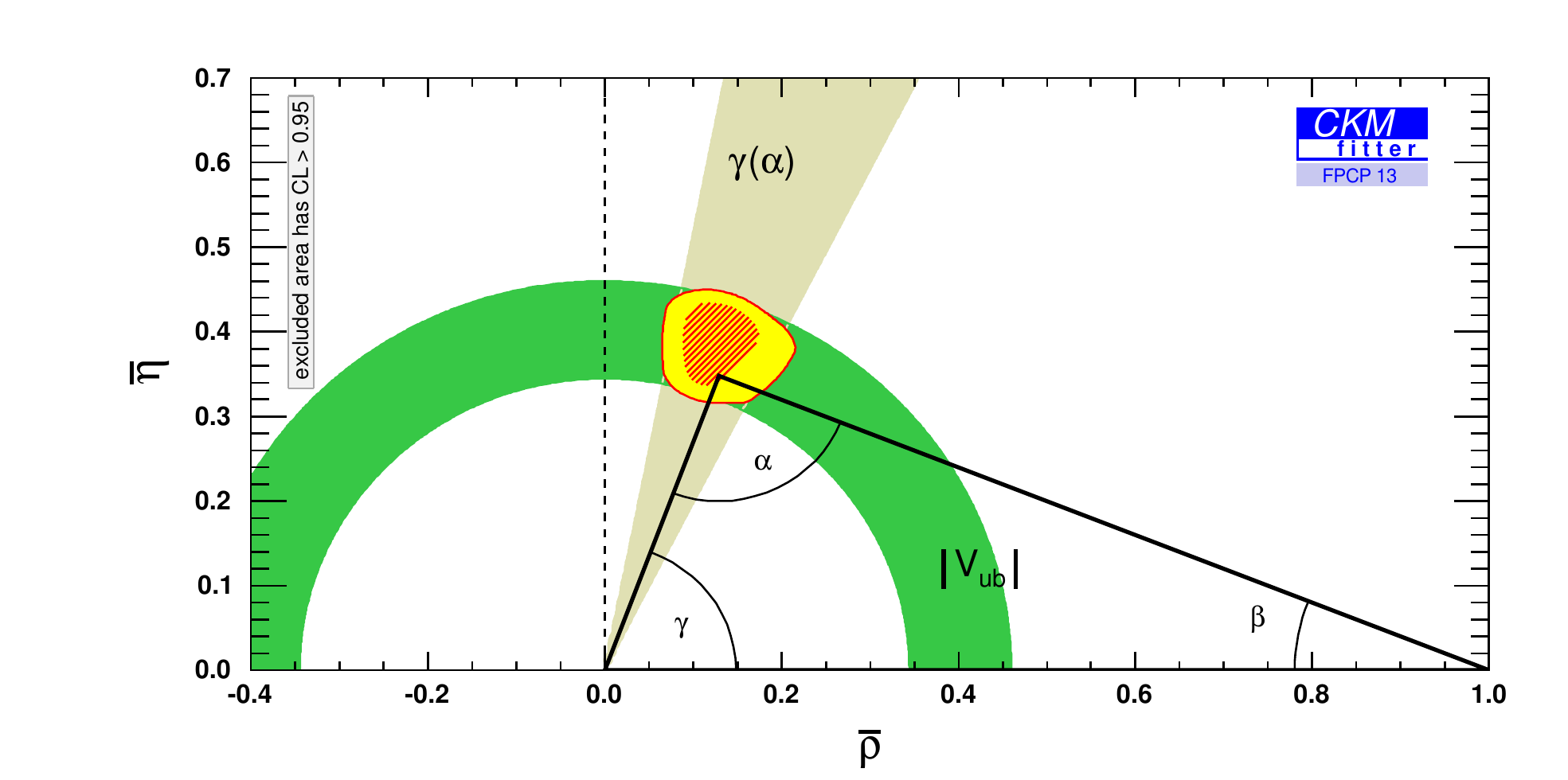}
\includegraphics[height=1.4in]{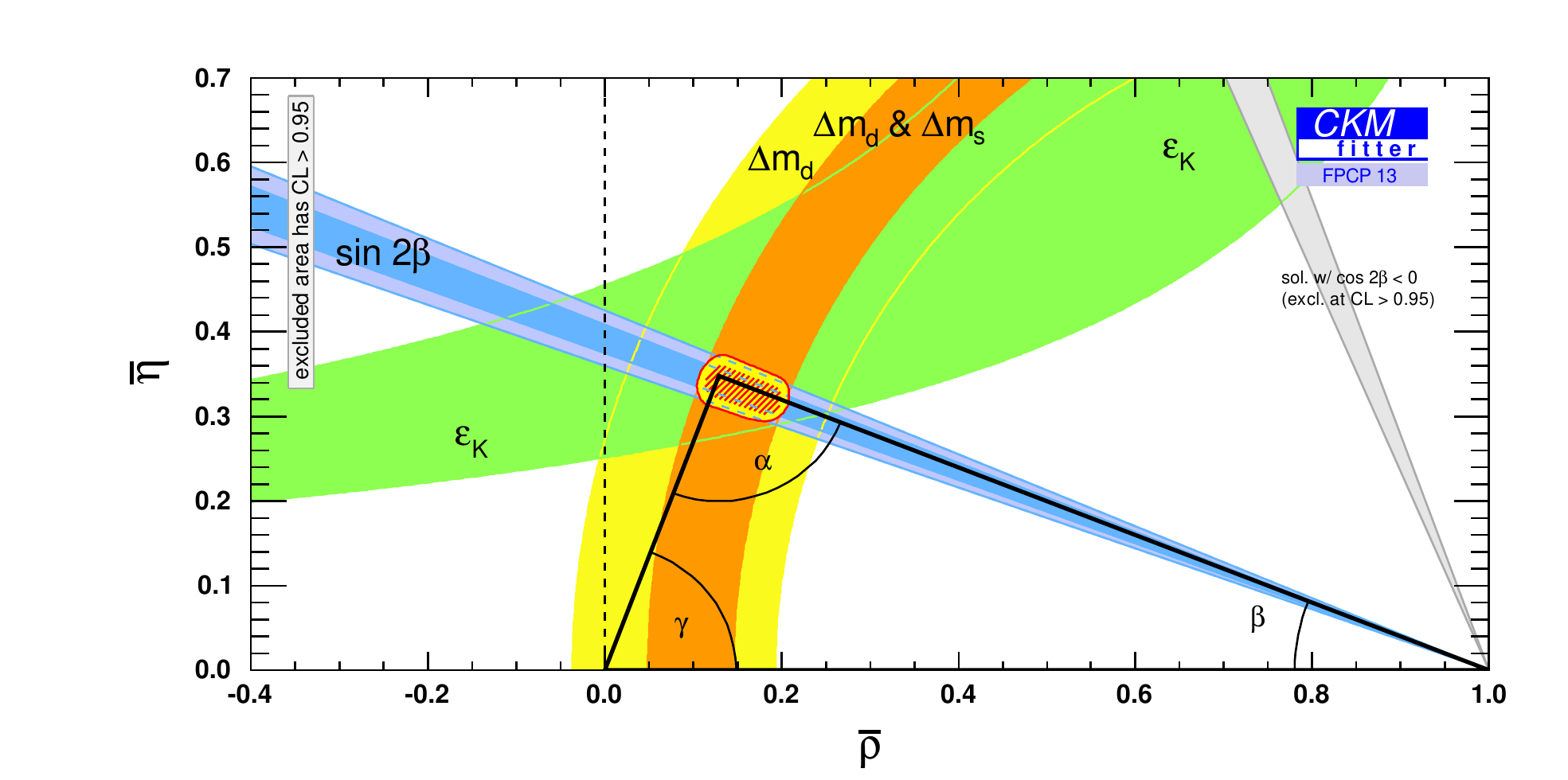}
\caption{Constraints on the CKM Unitarity Triangle due to (left) tree-level processes and (right) loop-mediated processes.}
\label{fig:CKMTriangle}
\end{figure}

Tree-level processes such as $B^\pm \rightarrow D K^\pm$ provide a theoretically clean measurement of $\gamma$ with no contributions from new physics processes. This measurement can be 
compared with measurements from loop-mediated processes, which are sensitive to new physics, to provide a test of the Standard Model. The current limits on the CKM Unitarity Triangle due to 
tree-level and loop processes, as calculated by the CKMFitter group~\cite{Charles:2004jd}, are shown in Fig.~\ref{fig:CKMTriangle}.

\subsection{GLW/ADS analysis of $B^\pm \rightarrow D K^\pm$ and $B^\pm \rightarrow D \pi^\pm$}
\label{sec:gamma:glwads}

The GLW method~\cite{Gronau:1991dp} uses $D$ decays to $C\!P$ eigenstates such as $K^+K^-$ and $\pi^+\pi^-$. Decays can proceed either via a $D^0$ or a $\bar{D}^0$ 
with a phase difference of $\delta_B + \gamma$. Suppression in the decay via $D^0$ with respect to the $\bar{D}^0$ decay limits interference to $\mathcal{O}(10\,\%)$ in $B^\pm \rightarrow D K^\pm$ and $\mathcal{O}(1\,\%)$ in $B^\pm \rightarrow D \pi^\pm$.

The ADS method~\cite{Atwood:1996ci} uses $D$ decays to quasi-flavour-specific states such as $\pi^+K^-$ and $\pi^-K^+\pi^+\pi^-$. 
Here the suppression of one of the $B$ decays is partially balanced by the suppression of one of the
$D$ decays, giving larger interference terms while also introducing an additional phase shift of $\delta_D$.

\begin{figure}[htb]
\centering
\includegraphics[height=2.in]{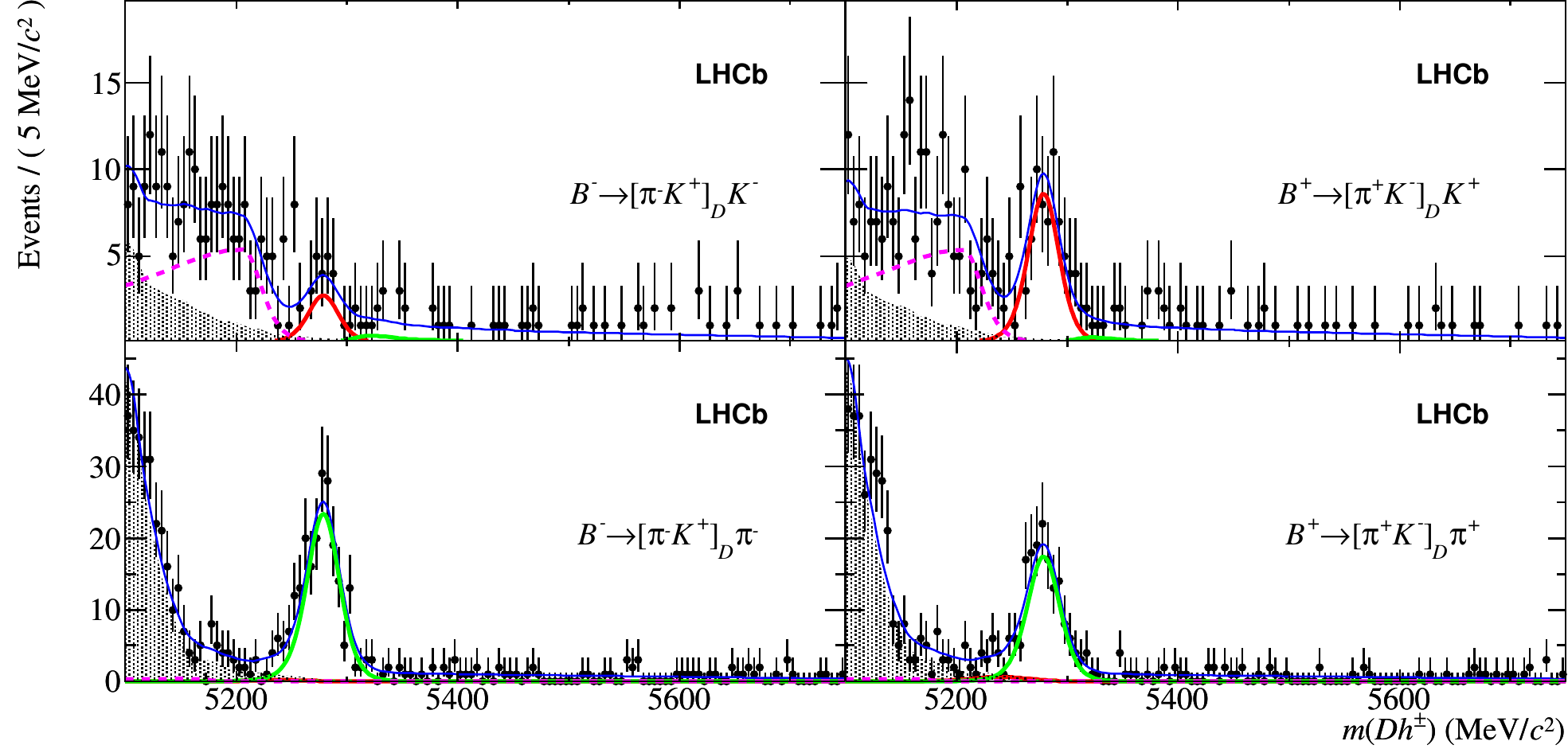}
\caption{Fits to the invariant mass distributions of the two-body suppressed ADS mode $\pi^\mp K^\pm$ in (top) $B^\mp \rightarrow D K^\mp$ and (bottom) $B^\mp \rightarrow D \pi^\mp$. The $B^\mp \rightarrow D K^\mp$ and $B^\mp \rightarrow D \pi^\mp$ components are shown in red and green, respectively. The shaded component indicates partially reconstructed backgrond, the dashed magenta line corresponds to partially reconstructed $\Lambda_b^0 \rightarrow \Lambda_c^+ h^-$ and the total shape also includes a combinatoric background.}
\label{fig:KpiMassFits}
\end{figure}

\begin{figure}[htb]
\centering
\includegraphics[height=2.in]{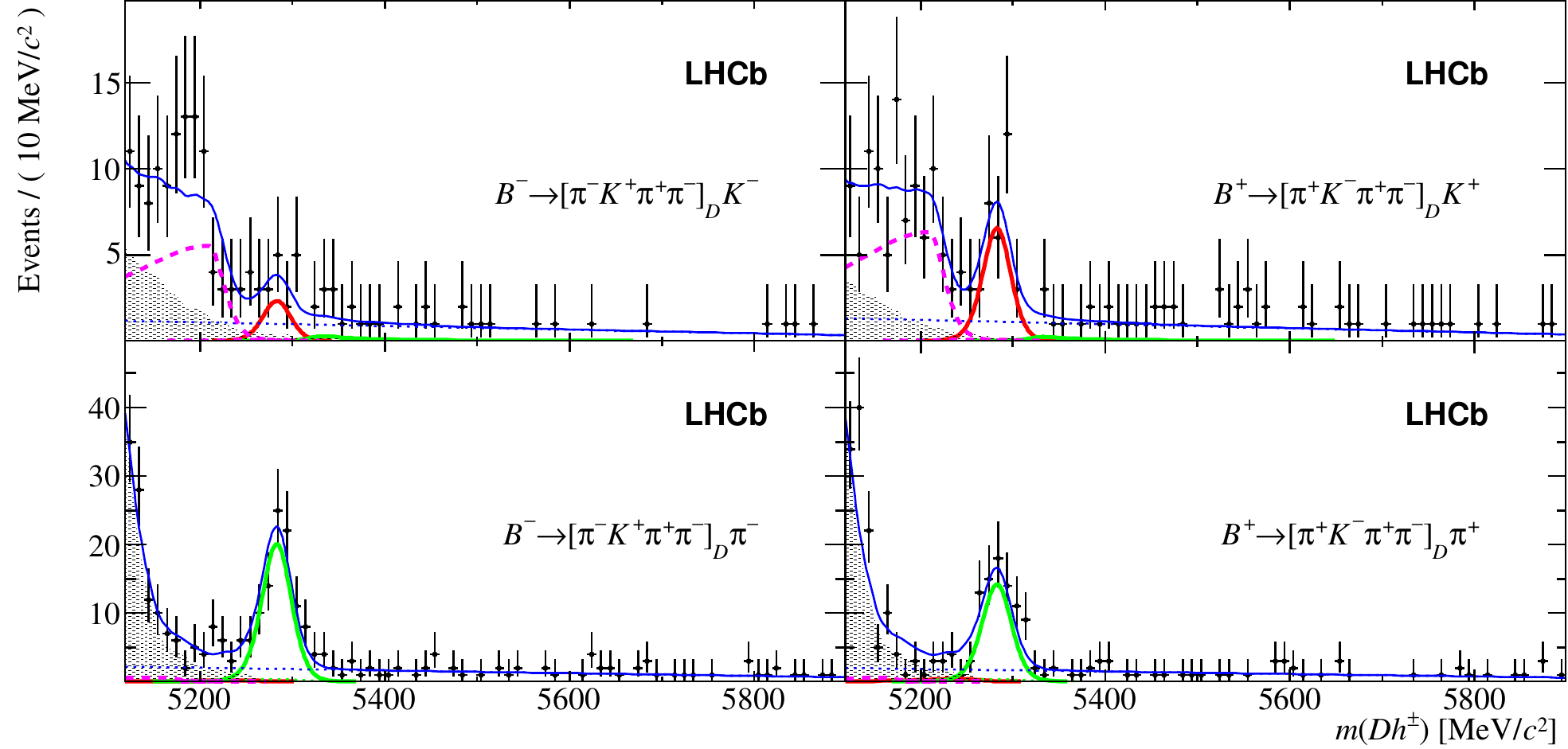}
\caption{Fits to the invariant mass distributions of the four-body suppressed ADS mode $\pi^\mp K^\pm\pi^\pm\pi^\mp$ in (top) $B^\mp \rightarrow D K^\mp$ and (bottom) $B^\mp \rightarrow D \pi^\mp$. The $B^\mp \rightarrow D K^\mp$ and $B^\mp \rightarrow D \pi^\mp$ components are shown in red and green, respectively. The shaded component indicates partially reconstructed backgrond, the dashed magenta line corresponds to partially reconstructed $B_s^0 \rightarrow DK^-\pi^+$ and the total shape also includes a combinatoric background.}
\label{fig:K3piMassFits}
\end{figure}

Analyses have been performed on $B^\pm \rightarrow D K^\pm$ and $B^\pm \rightarrow D \pi^\pm$ with the $D$ meson reconstructed from the final states $K^+K^-$, $\pi^+\pi^-$, $K^+\pi^-$,
$\pi^+K^-$, $K^-\pi^+\pi^+\pi^-$ and $\pi^-K^+\pi^+\pi^-$ using LHCb data corresponding to $1\,\text{fb}^{-1}$ of $pp$ collisions 
at a centre of mass energy of 7 TeV~\cite{Aaij:2012kz,Aaij:2013mba}. 
The invariant mass distributions of the two- and four-body suppressed ADS modes are shown in Fig.~\ref{fig:KpiMassFits} and Fig.~\ref{fig:K3piMassFits}, respectively.
The observables measured are the ratio of $DK$ to $D\pi$ for each $D$ final state,
\begin{equation}
R_{K/\pi}^f = \frac{\Gamma\left(B^- \rightarrow D\left[\rightarrow f\right]K^-\right)+\Gamma\left(B^+ \rightarrow D\left[\rightarrow \bar{f}\right]K^+\right)}{\Gamma\left(B^- \rightarrow D\left[\rightarrow f\right]\pi^-\right)+\Gamma\left(B^+ \rightarrow D\left[\rightarrow \bar{f}\right]\pi^+\right)} \, ,\nonumber
\end{equation}
the charge asymmetry for each final state,
\begin{equation}
A_h^f = \frac{\Gamma\left(B^- \rightarrow D\left[\rightarrow f\right]h^-\right) - \Gamma\left(B^+ \rightarrow D\left[\rightarrow \bar{f}\right]h^+\right)}{\Gamma\left(B^- \rightarrow D\left[\rightarrow f\right]h^-\right) + \Gamma\left(B^+ \rightarrow D\left[\rightarrow \bar{f}\right]h^+\right)} \, ,\nonumber
\end{equation}
and the ratio of the suppressed to favoured modes for $D \rightarrow K\pi$ and $D \rightarrow K\pi\pi\pi$,
\begin{equation}
R_h^\pm = \frac{B^\pm \rightarrow D\left[f_\text{sup}\right]h^\pm}{B^\pm \rightarrow D\left[f\right]h^\pm} \, .\nonumber
\end{equation}
The values obtained for each of these observables can be found in Refs.~\cite{Aaij:2012kz,Aaij:2013mba}. These variables serve as inputs for the combined $\gamma$ measurements in Section~\ref{sec:gamma:comb1fb-1} and Section~\ref{sec:gamma:comb3fb-1}.

\subsection{GGSZ analysis of $B^\pm \rightarrow D K^\pm$}
\label{sec:gamma:ggsz}

\begin{figure}[htb]
\centering
\includegraphics[height=2.in]{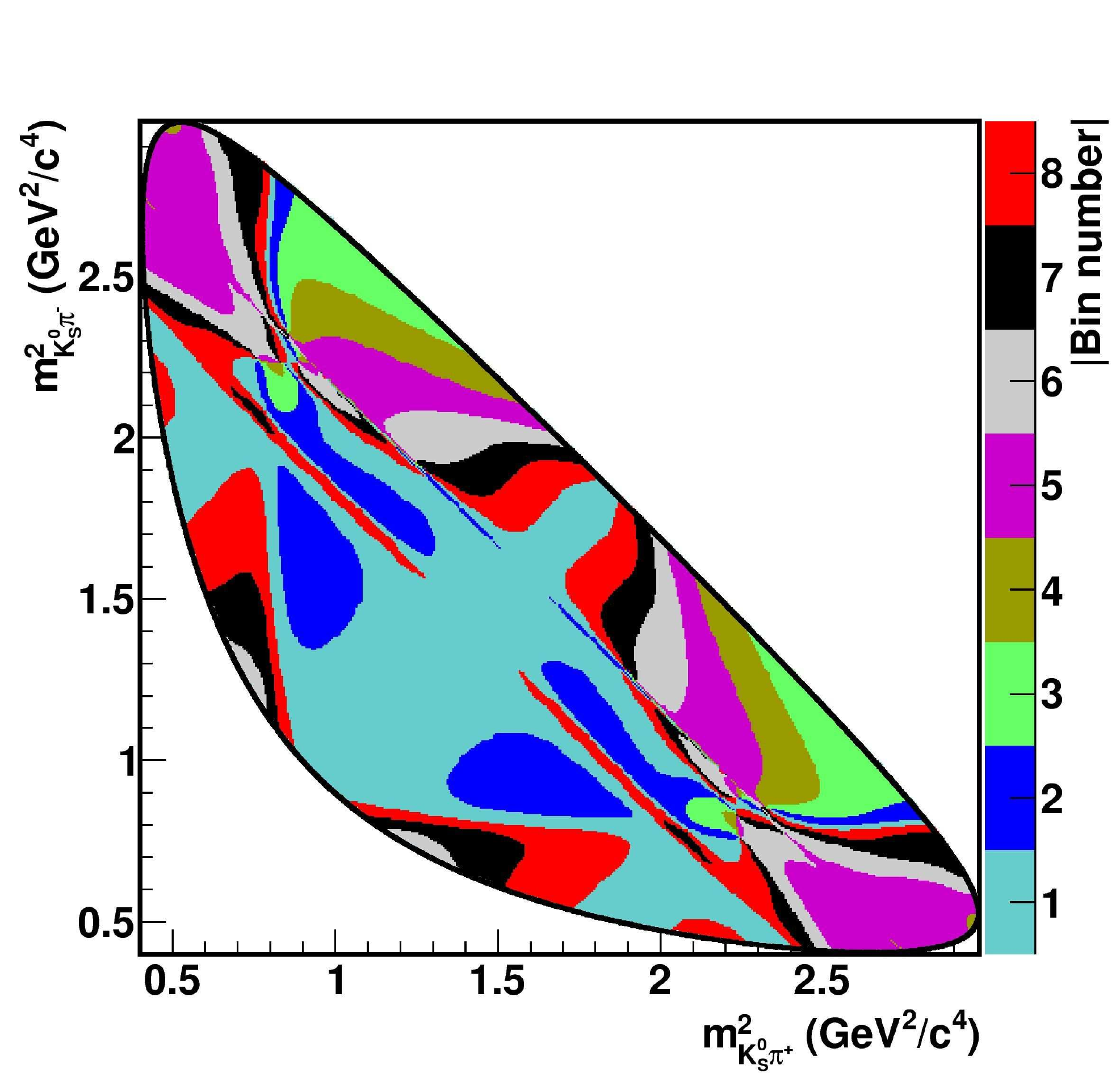}
\includegraphics[height=2.in]{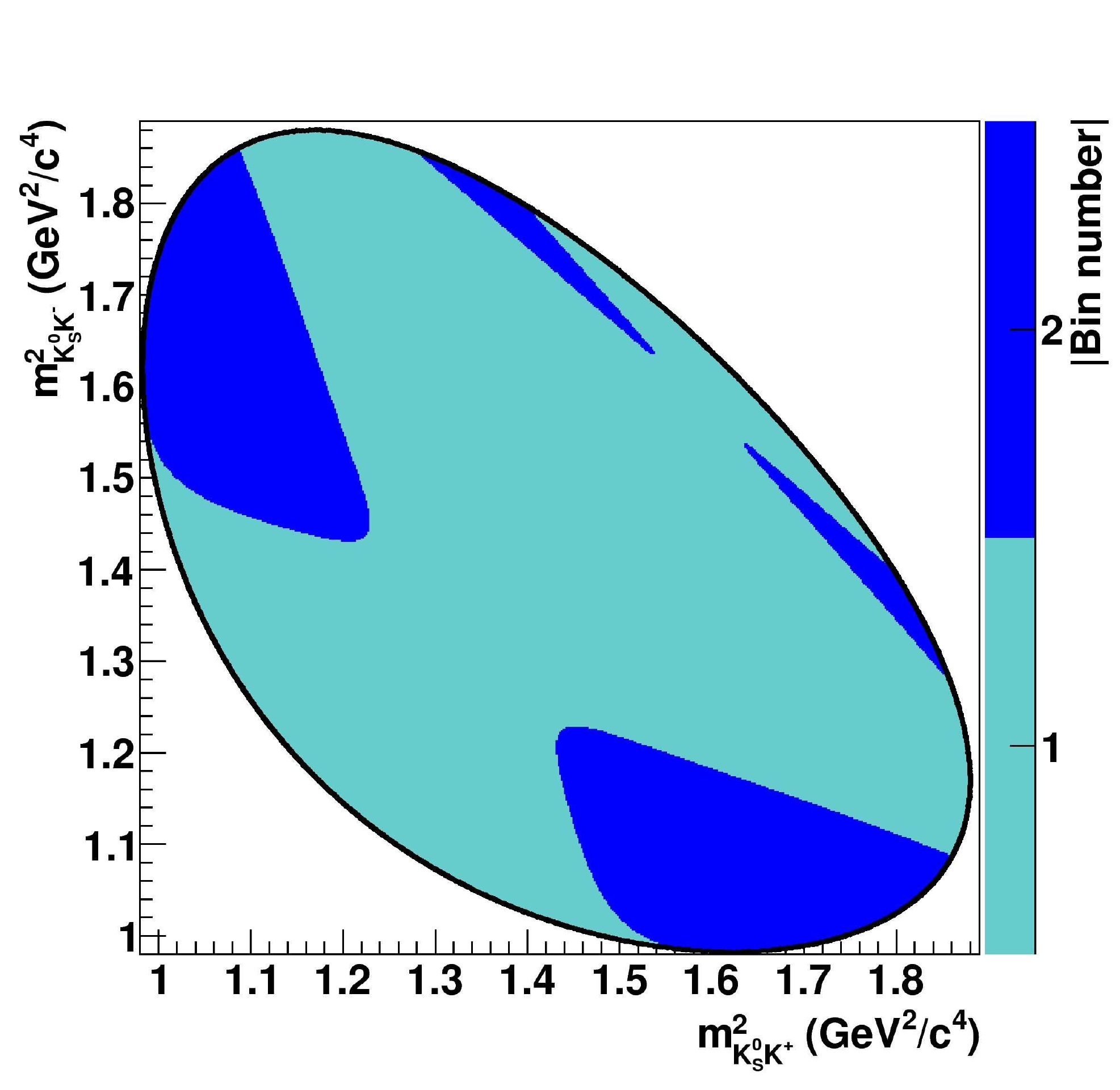}
\caption{Binning schemes used for the Dalitz plots of (left) $D \rightarrow K_S^0 \pi^+ \pi^-$ and (right) $D \rightarrow K_S^0 K^+ K^-$. Bins in the top-left half of the plots ($m^2_{K_S^0 h^-}>m^2_{K_S^0 h^+}$) are identified as $+i$ and bins in the bottom-right half are labeled $-i$.}
\label{fig:ggszDDalitz}
\end{figure}

The GGSZ method~\cite{Giri:2003ty} exploits the variation of the strong phase $\delta_D$ across the Dalitz plot in $D$ decays to three-body self-conjugate states 
such as $K_S^0 \pi^+ \pi^-$ and $K_S^0 K^+ K^-$. The Dalitz plot is divided into bins, as shown in Fig.~\ref{fig:ggszDDalitz}, chosen to maximise statistical sensitivity. 
The populations of $B^+$ and $B^-$ decays in each bin are given by 

\begin{equation}
N^+_{\pm i} = h_{B^+}\left[K_{\mp i} + (x^2_+ + y^2_+)K_{\pm i} + 2 \sqrt{K_iK_{-i}}(x_+c_{\pm i} \mp y_+ s_{\pm i})\right] \, ,\nonumber
\end{equation}
\begin{equation}
N^-_{\pm i} = h_{B^-}\left[K_{\pm i} + (x^2_- + y^2_-)K_{\mp i} + 2 \sqrt{K_iK_{-i}}(x_-c_{\pm i} \pm y_- s_{\pm i})\right] \, , \nonumber
\end{equation}
where $K_{\pm i}$ is the efficiency corrected yield in bin $\pm i$ due to $D^0$ flavour tagged events from BaBar~\cite{Aubert:2008bd,delAmoSanchez:2010rq} 
and $c_{\pm i}$ and $s_{\pm i}$ are the cosine and sine of the strong phase $\delta_D$ in bin $\pm i$ from CLEO-c~\cite{Libby:2010nu}.

The remaining parameters are left free in  the fit to the data: $h_{B^\pm}$ are normalisation factors for $B^\pm$, and $x_\pm = r_B \text{cos}(\delta_B \pm \gamma)$ and 
$y_\pm = r_B \text{sin}(\delta_B \pm \gamma)$ are the Cartesian parameters, which are sensitive to $\gamma$.

Analyses have been performed on $B^\pm \rightarrow D K^\pm$ with the $D$ meson reconstructed in the final states $K_S^0 \pi^+ \pi^-$ and $K_S^0 K^+ K^-$ 
using LHCb data corresponding to $1\,\text{fb}^{-1}$ of $pp$ collisions at a centre of mass energy of 7 TeV~\cite{Aaij:2012hu} and $2\,\text{fb}^{-1}$ of $pp$ collisions at a centre of mass energy of 8 TeV~\cite{LHCb-CONF-2013-004}.
The values obtained for the Cartesian parameters in the 8 TeV analysis are
\begin{equation}
x_+ = \left( -8.7 \pm 3.1 \text{(stat.)} \pm 1.6 \text{(syst.)} \pm 0.6 \text{(ext.)} \right) \times10^{-2} \, ,\nonumber
\end{equation}
\begin{equation}
x_- = \left( \phantom{-}5.3 \pm 3.2 \text{(stat.)} \pm 0.9 \text{(syst.)} \pm 0.9 \text{(ext.)} \right) \times10^{-2} \, ,\nonumber
\end{equation}
\begin{equation}
y_+ = \left( \phantom{-}0.1 \pm 3.6 \text{(stat.)} \pm 1.4 \text{(syst.)} \pm 1.9 \text{(ext.)} \right) \times10^{-2} \, ,\nonumber
\end{equation}
\begin{equation}
y_- = \left( \phantom{-}9.9 \pm 3.6 \text{(stat.)} \pm 2.2 \text{(syst.)} \pm 1.6 \text{(ext.)} \right) \times10^{-2} \, ,\nonumber
\end{equation}
where the third uncertainty is due to the CLEO-c strong phase measurements used in the fit.

\begin{figure}[htb]
\centering
\includegraphics[height=2.in]{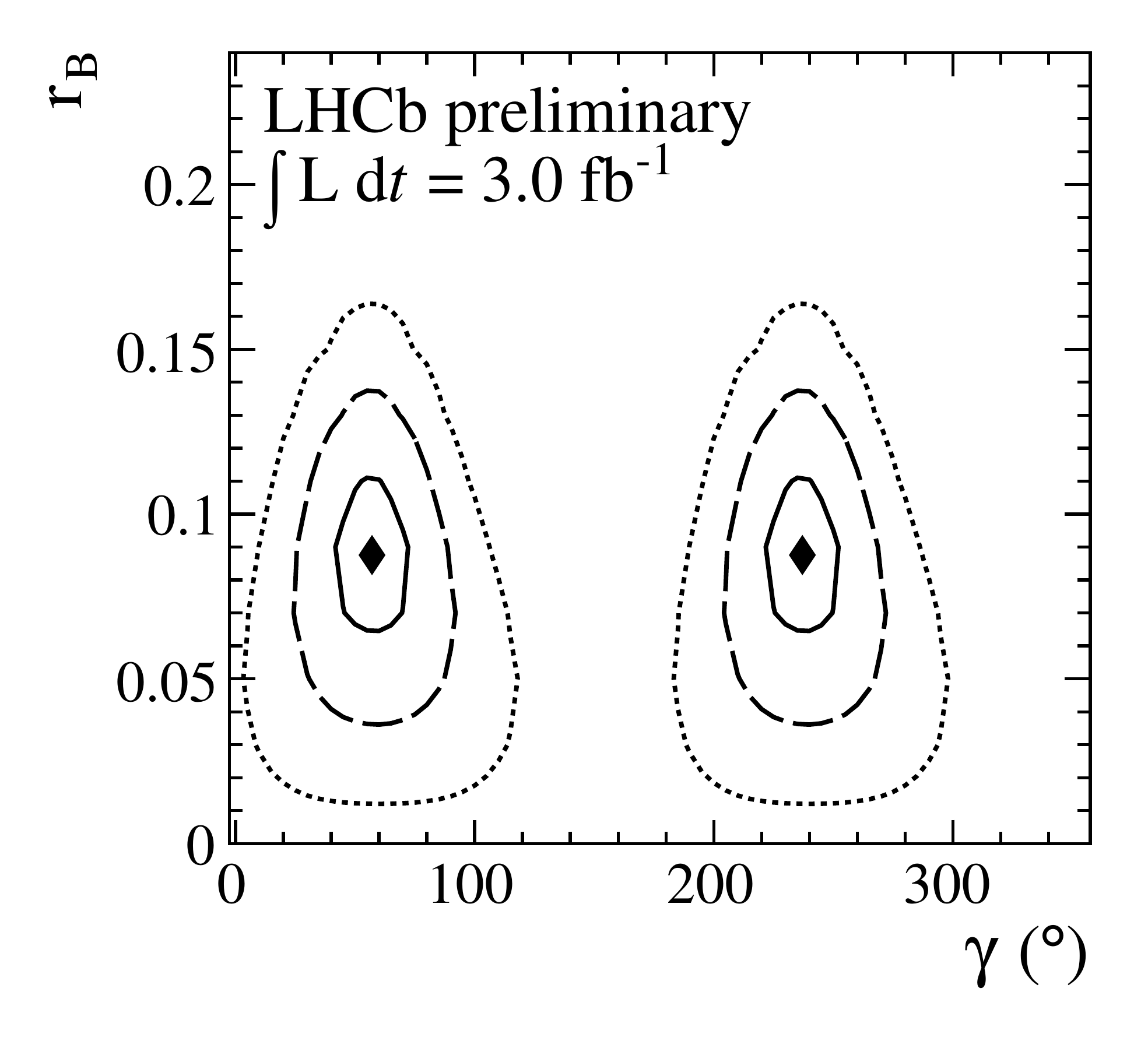}
\includegraphics[height=2.in]{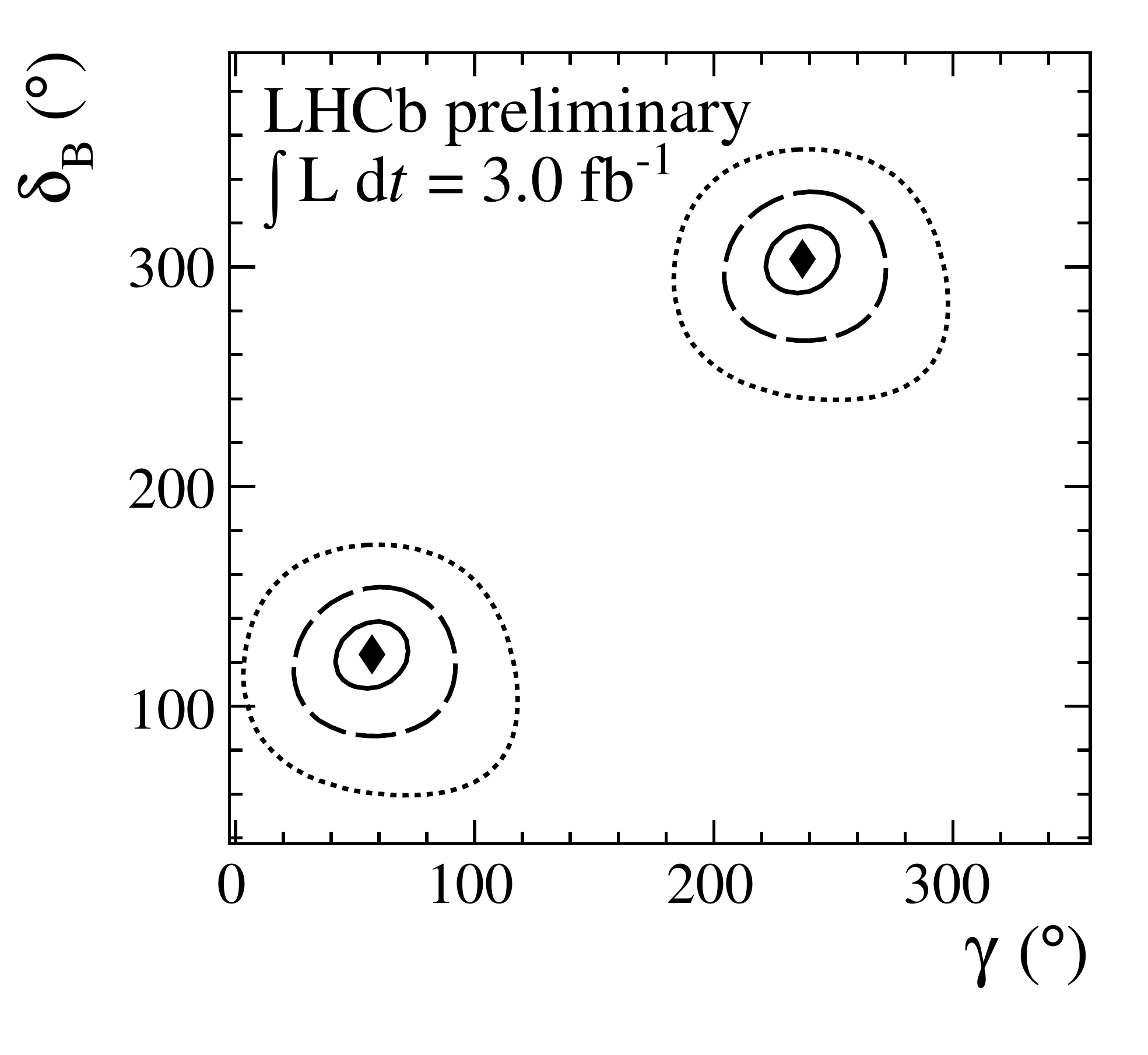}
\caption{Two-dimensional projections of the confidence regions onto the (left) $(\gamma, r_B)$ and (right) $(\gamma, \delta_B)$ planes. Contours indicate the $1$, $2$ and $3\sigma$ boundaries and diamonds mark the central values.}
\label{fig:ggsz2DCR}
\end{figure}

Combining these values with the results from the 7 TeV analysis  and fitting for $\gamma$, $r_B$ and $\delta_B$ yields values of $(57 \pm 16)^\circ$, $(8.8^{+2.3}_{-2.4})\times10^{-2}$ and $(124^{+15}_{-17})^\circ$, respectively, where the values for $\gamma$ and $\delta_B$ are modulo $180^\circ$. 
Two-dimensional projections of the confidence regions for these parameters are shown in Fig.~\ref{fig:ggsz2DCR}.

\subsection{Combination of results from $1\,\text{fb}^{-1}$ measurements}
\label{sec:gamma:comb1fb-1}

The results in Section~\ref{sec:gamma:glwads} and Section~\ref{sec:gamma:ggsz} are combined using a frequentist approach to obtain a more constraining measurement of $\gamma$ ~\cite{Aaij:2013zfa}. 
In addition to these results further measurements are included to improve the fit: measurements of the strong phases and coherence factors for 
$D \rightarrow K\pi$ and $D \rightarrow K\pi\pi\pi$ decays from CLEO-c\cite{Lowery:2009id}, 
$C\!P$ asymmetry measurements of the neutral $D$ mesons from the Heavy Flavour Averaging Group\cite{Amhis:2012bh} and charm mixing parameters from LHCb\cite{Aaij:2012nva}.
A likelihood is constructed from the measured observables as
\begin{equation}
{\cal L}\left(\vec{\alpha}\right) = {\displaystyle\prod_i} \xi_i\left(\vec{A}_i^{\text{obs}}|\vec{\alpha}\right) \, ,\nonumber
\end{equation}
where the sum is over the different measurements, $\vec{\alpha}$ is the set of parameters and $\xi_i$ denotes the likelihood probability density functions (PDFs) of the observables $\vec{A}_i^{\text{obs}}$. For most observables a Gaussian PDF is assumed, however, where highly non-Gaussian behaviour is observed, the experimental likelihood is used.

\begin{figure}[htb]
\centering
\includegraphics[height=1.3in]{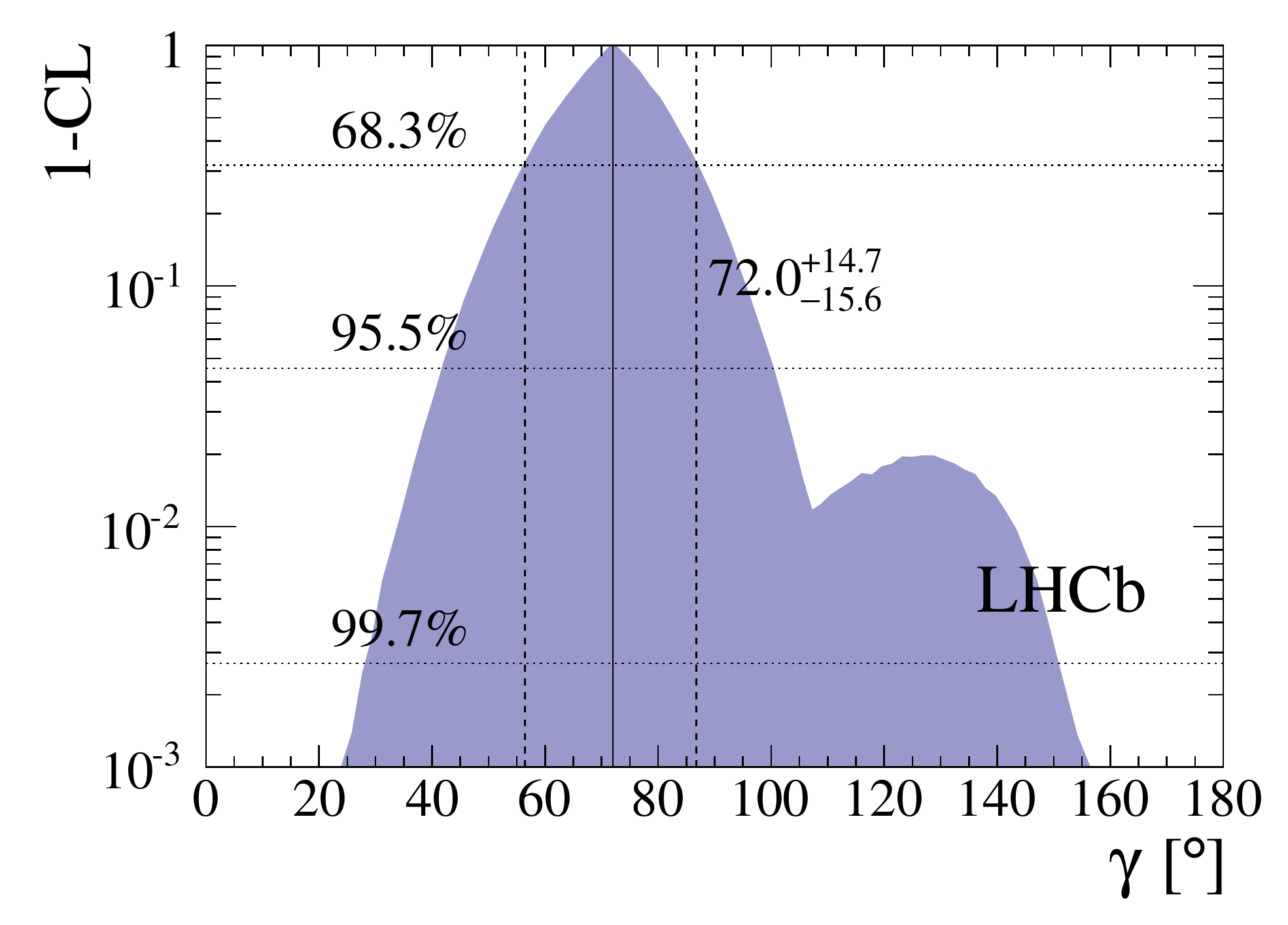}
\includegraphics[height=1.3in]{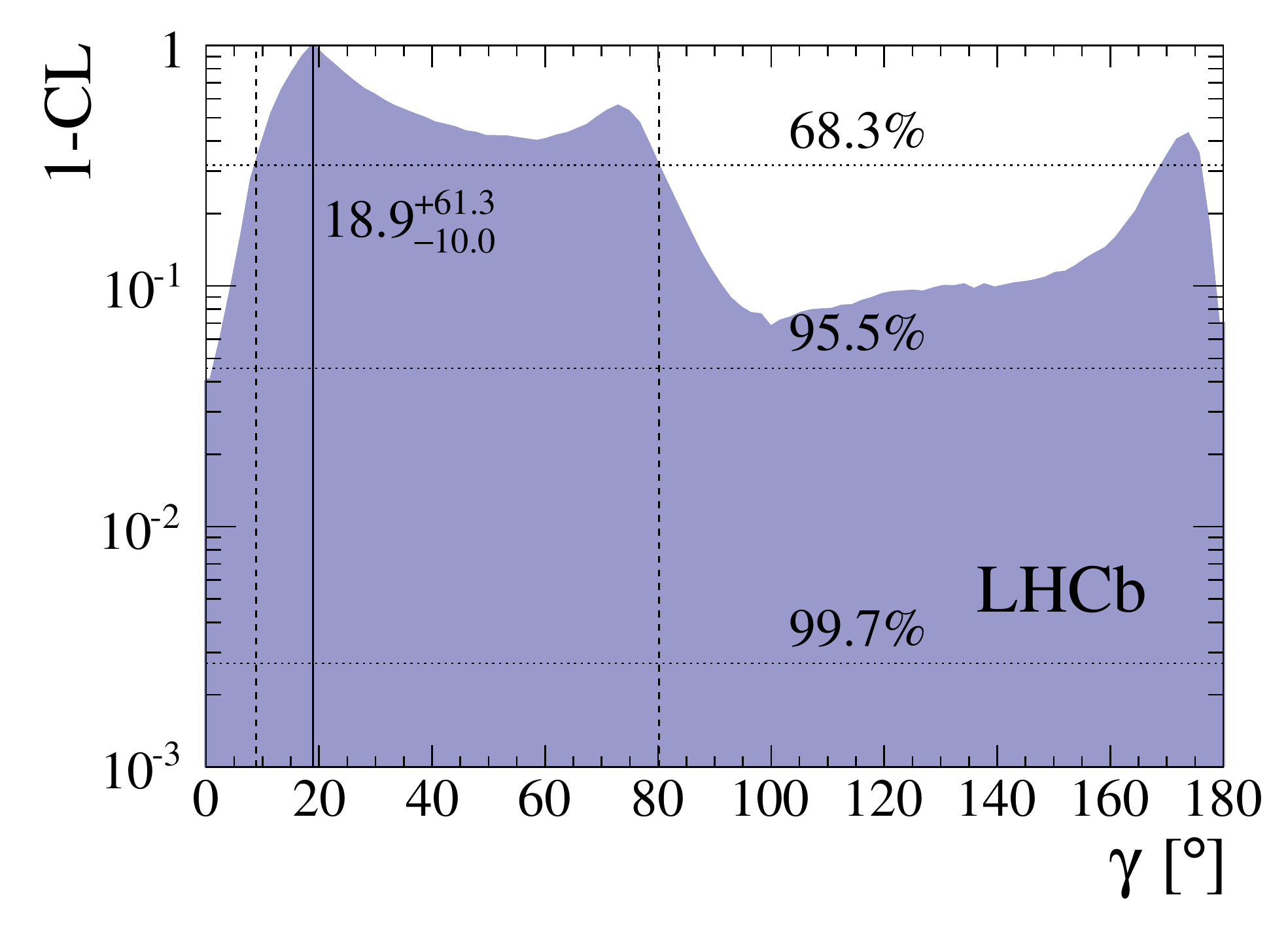}
\includegraphics[height=1.3in]{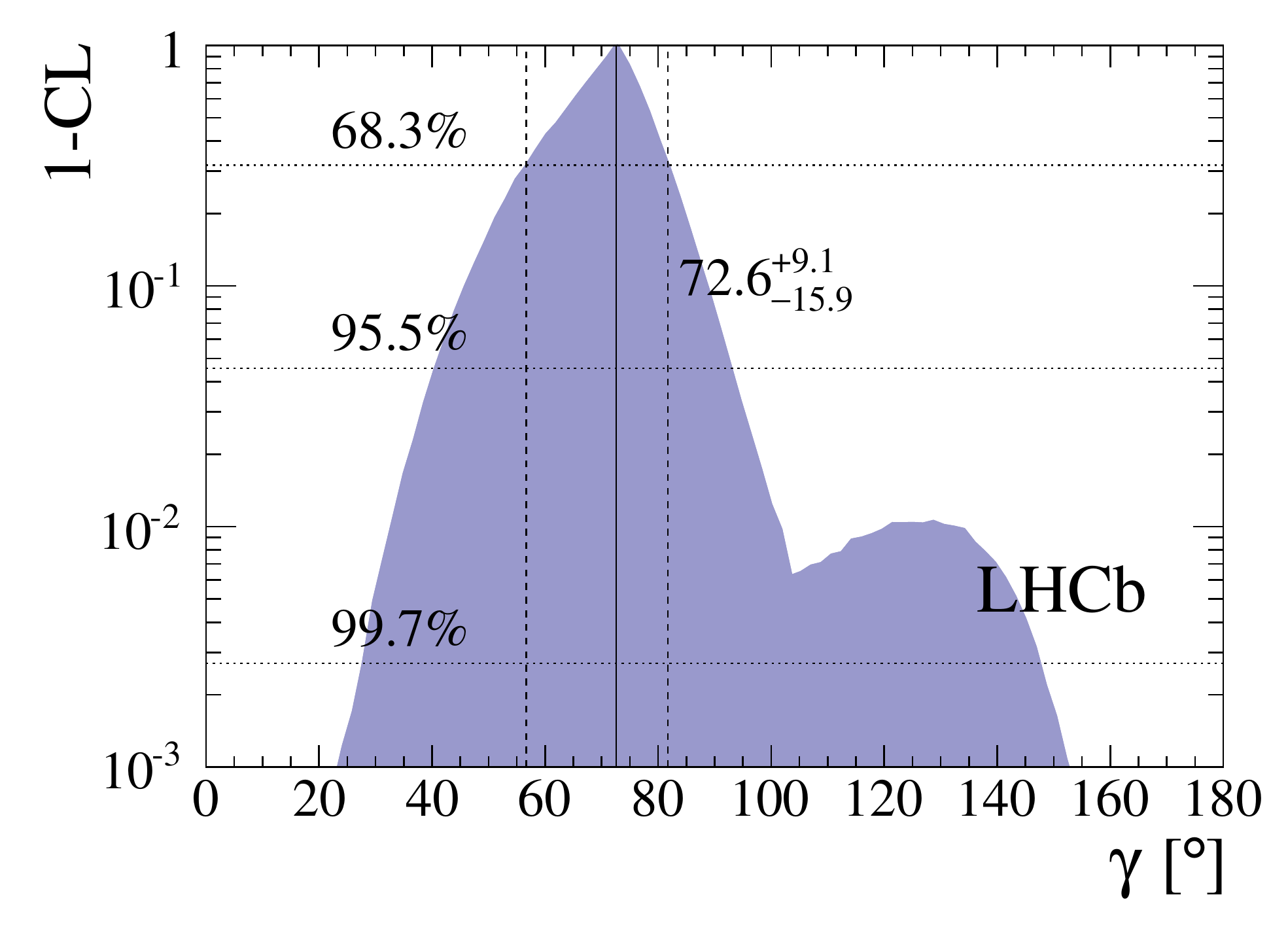}
\caption{$1-\text{CL}$ curves for $\gamma$ from the combined $1\,\text{fb}^{-1}$ GLW/ADS and $1\,\text{fb}^{-1}$ GGSZ measurements using (left) only $DK$, (centre) only $D\pi$ and (right) both decay modes.}
\label{fig:comb1gammaCL}
\end{figure}

A combined $\gamma$ measurement has been performed including the results from Section~\ref{sec:gamma:glwads} and a subset of the results from Section~\ref{sec:gamma:ggsz} corresponding to $1\,\text{fb}^{-1}$ of $pp$ collisions 
at a centre of mass energy of 7 TeV~\cite{Aaij:2012hu}.
The best-fit values and confidence intervals (modulo $180^\circ$) of $\gamma$ are given in Table~\ref{tab:comb1} and the $1-\text{CL}$ curves for $\gamma$ are shown in Fig.~\ref{fig:comb1gammaCL}. 

\begin{table}[t]
\begin{center}
\begin{tabular}{lccc}  
combination &  $\gamma$ & 68\,\% CL & 95\,\% CL\\ \hline
 $DK$  &   $72.0^\circ$ & $[56.4, 86.7]^\circ$ & $[42.6, 99.6]^\circ$ \\
 $D\pi$  &   $18.9^\circ$ & $[7.4, 99.2]^\circ \cup [167.9, 176.4]^\circ$ & - \\
 $DK$ and $D\pi$  &   $72.6^\circ$ & $[55.4, 82.3]^\circ$ & $[40.2, 92.7]^\circ$ \\
\end{tabular}
\caption{Best-fit values and confidence intervals for $\gamma$ from the combination of $DK$ and $D\pi$ measurements.}
\label{tab:comb1}
\end{center}
\end{table}

\subsection{Combination including $3\,\text{fb}^{-1}$ GGSZ measurement}
\label{sec:gamma:comb3fb-1}

Another combination~\cite{LHCb-CONF-2013-006} has been performed that incorporates all of the results reported in Section~\ref{sec:gamma:ggsz} but only those observables from Section~\ref{sec:gamma:glwads} corresponding to $B^\pm \rightarrow DK^\pm$ decays.
Mixing in the neutral $D$ mesons is also neglected in the equations used for the observables in this combination.

\begin{figure}[htb]
\centering
\includegraphics[height=2.in]{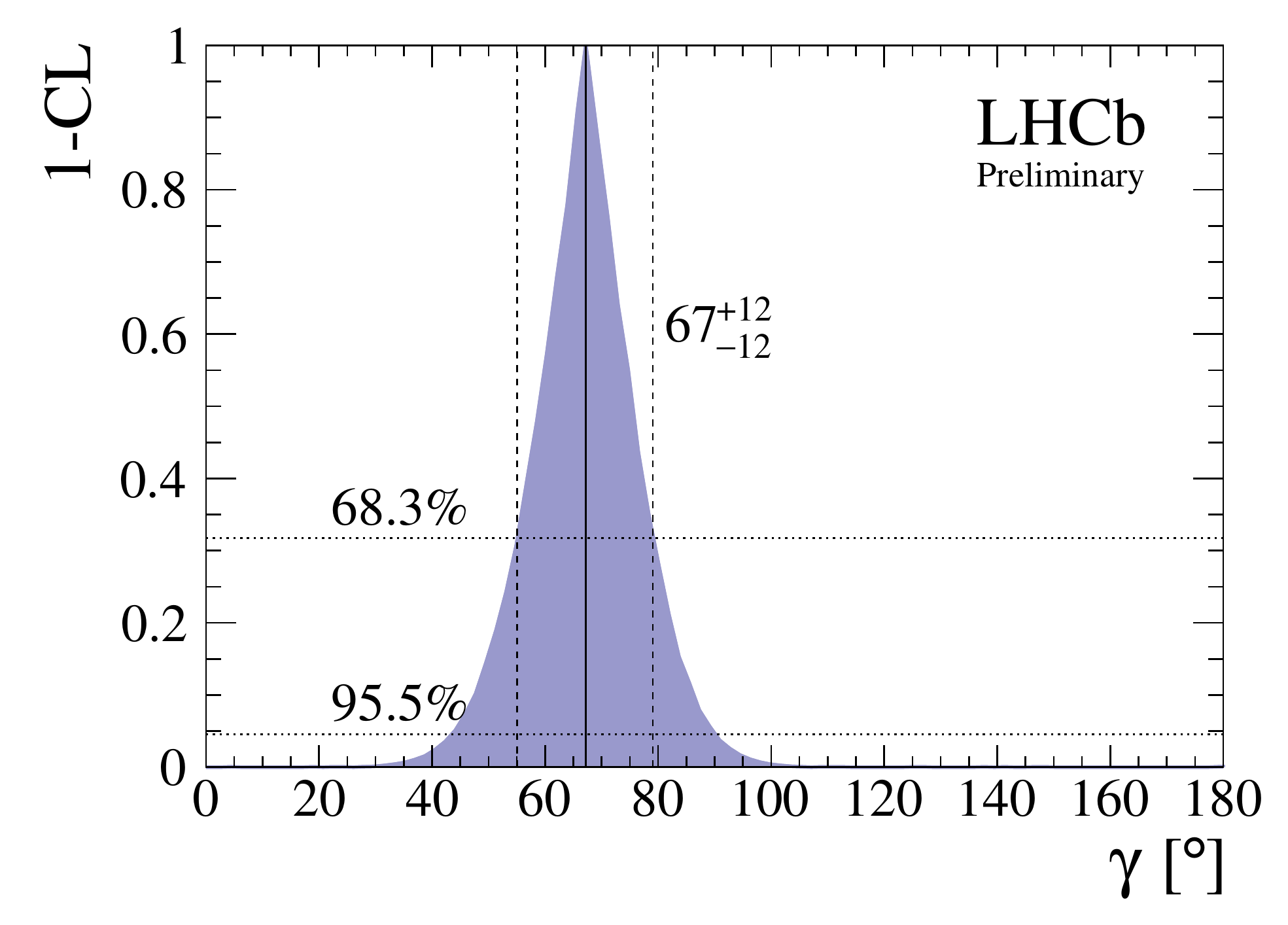}
\caption{$1-\text{CL}$ curve for $\gamma$ from the combined $1\,\text{fb}^{-1}$ GLW/ADS and $3\,\text{fb}^{-1}$ GGSZ measurements.}
\label{fig:comb2gammaCL}
\end{figure}

\begin{figure}[htb]
\centering
\includegraphics[height=2.in]{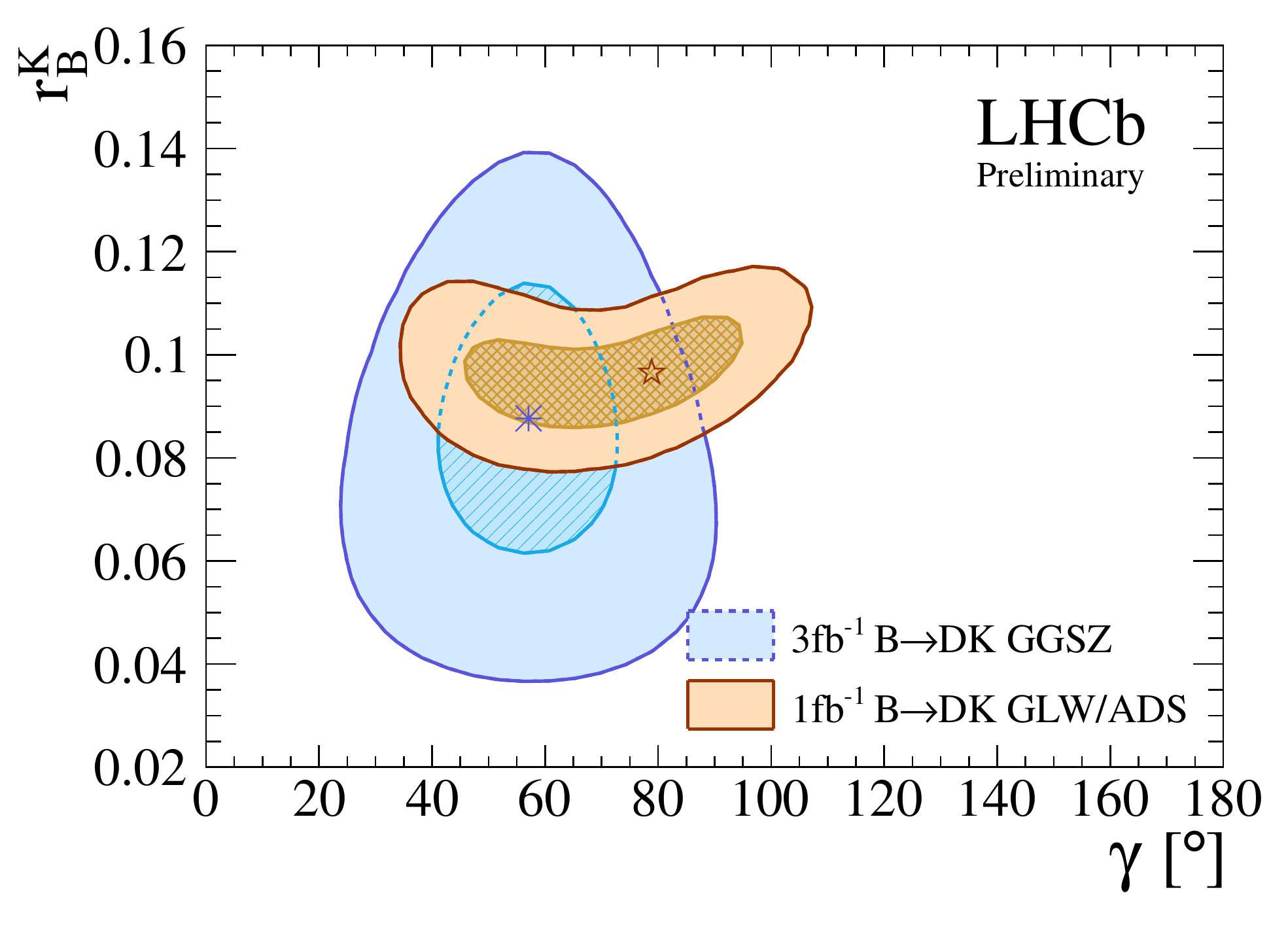}
\caption{Two-dimensional projection of the confidence regions onto the $(\gamma, r_B)$ plane. Contours show the $1$ and $2\sigma$ boundaries and markers indicate the central values.}
\label{fig:comb2gammaRB}
\end{figure}

The best-fit values and confidence intervals (all modulo $180^\circ$) for $\gamma$, $r_B$ and $\delta_B$ are given in Table~\ref{tab:comb2}. Figure~\ref{fig:comb2gammaCL} and Figure~\ref{fig:comb2gammaRB} show the $1-\text{CL}$ curve for $\gamma$, and the 2D projection of the likelihood in $\gamma$ and $r_B$, respectively.

\begin{table}[t]
\begin{center}
\begin{tabular}{lccc}  
quantity &  value & 68\,\% CL & 95\,\% CL\\ \hline
 $\gamma$  &   $67.2^\circ$ & $[55.1, 79.1]^\circ$ & $[43.9, 89.5]^\circ$ \\
 $r_B$  &   $0.0923$ & $[0.0843, 0.1001]$ & $[0.0762, 0.1075]$ \\
 $\delta_B$  &   $114.3^\circ$ & $[101.3, 126.3]^\circ$ & $[88.7, 136.3]^\circ$ \\
\end{tabular}
\caption{Best-fit values and confidence intervals for $\gamma$, $r_B$ and $\delta_B$ from the combination of $DK$ measurements including GGSZ measurements from $3\,\text{fb}^{-1}$ of data.}
\label{tab:comb2}
\end{center}
\end{table}

\section{Measurement of $B_{(s)}^0 \rightarrow DK\pi$ branching fractions}

The decay mode $B^0 \rightarrow D K^+ \pi^-$ has potential for a significant future measurement of $\gamma$~\cite{Gronau:2002mu,Gershon:2008pe,Gershon:2009qc}. 
Sensitivity to $\gamma$ comes from the interference of $b \rightarrow c$ and $b \rightarrow u$ amplitudes of a similar magnitude. 
$B_s^0 \rightarrow D K^- \pi^+$ and the related mode $B_s^0 \rightarrow D^* K^- \pi^+$ form important backgrounds to this mode, 
therefore, an understanding of these modes is necessary.

\begin{figure}[htb]
\centering
\includegraphics[height=2.in]{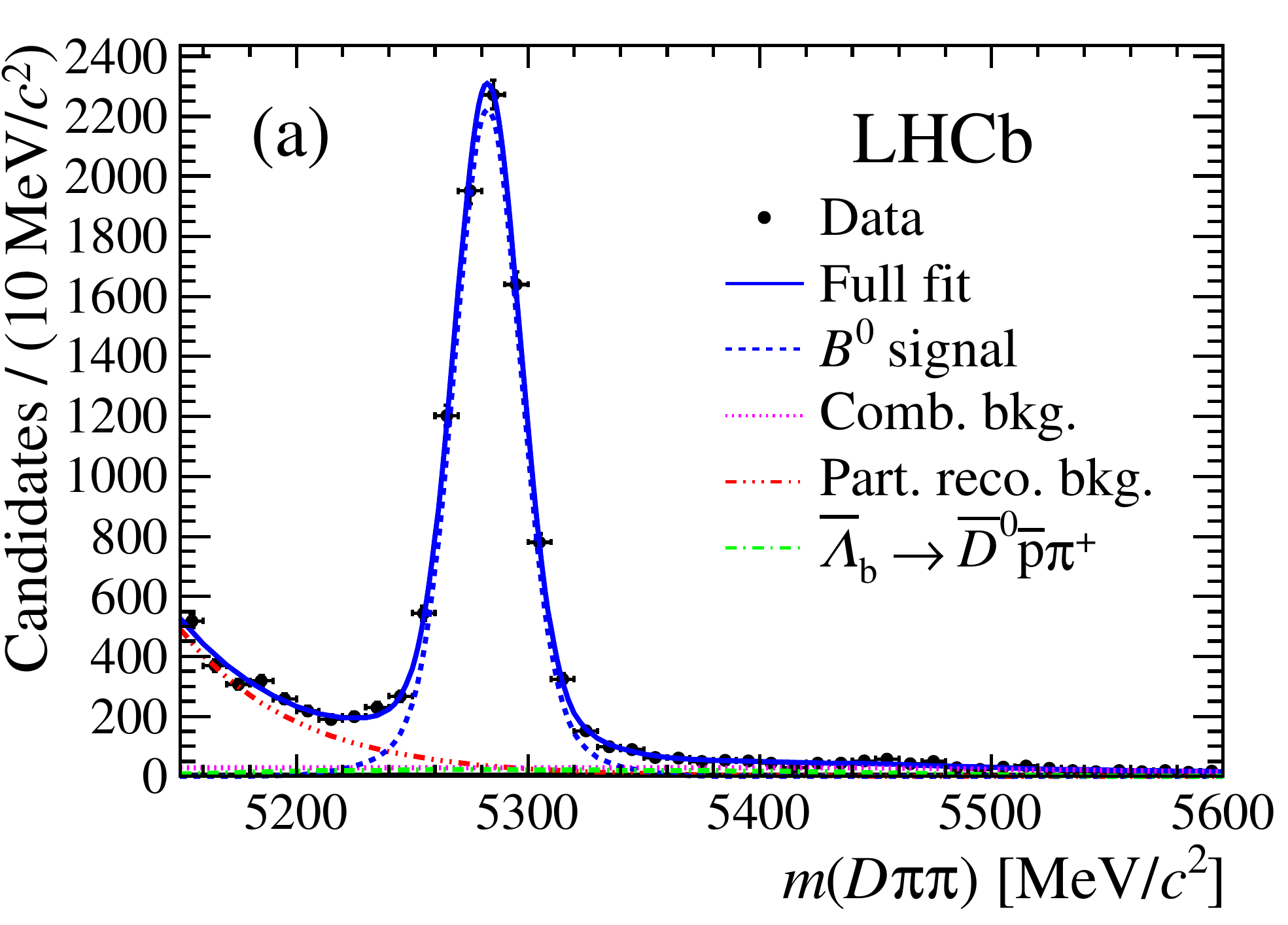}
\includegraphics[height=2.in]{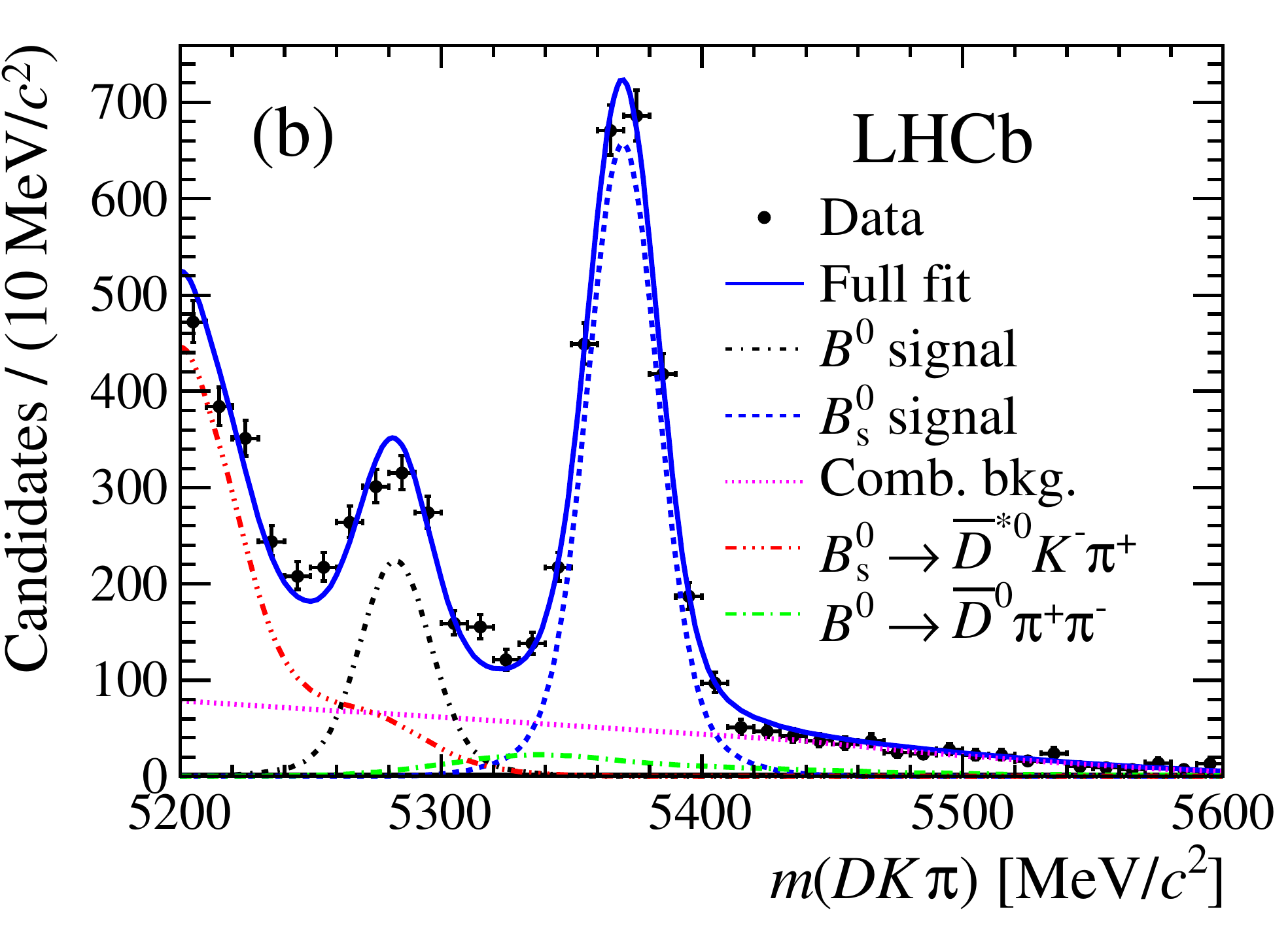}
\caption{Fits to the $B^0_{(s)}$ candidate invariant mass distributions for the (a)~$D\pi\pi$ and (b)~$DK\pi$ samples. 
    Data points are shown in black, the full fitted PDFs as solid blue lines and the components as detailed in the legends.}
\label{fig:dkpiFit}
\end{figure}

Branching fraction measurements of $B^0 \rightarrow D K^+ \pi^-$ and $B_s^0 \rightarrow D K^- \pi^+$, relative to the normalisation mode $B^0 \rightarrow D \pi^+ \pi^-$, 
have been made using LHCb data corresponding to $1\,\text{fb}^{-1}$ of $pp$ collisions at a centre of mass energy of 7 TeV~\cite{Aaij:2013pua}. 
The invariant mass distributions of $D\pi\pi$ and $DK\pi$ candidates where the $D$ is reconstructed from $\bar{D}^0 \rightarrow K^+ \pi^-$
are shown in Fig.~\ref{fig:dkpiFit}. The measured relative branching fractions are

\begin{equation}
\frac{
  {\cal B}\left(B^0 \to \bar{D}^0 K^+\pi^-\right)}{
  {\cal B}\left(B^0 \to \bar{D}^0 \pi^+\pi^-\right)} = 0.106 \pm 0.007\,\text{(stat.)} \pm 0.008\,\text{(syst.)} \, ,\nonumber
\end{equation} 
\begin{equation}
\frac{
  {\cal B}\left(B_s^0 \to \bar{D}^0 K^-\pi^+\right)}{
  {\cal B}\left(B^0 \to \bar{D}^0 \pi^+\pi^-\right)} = 1.18 \pm 0.05\,\text{(stat.)} \pm 0.12\,\text{(syst.)} \, .\nonumber
\end{equation}

These relative measurements yield absolute branching fractions of
\begin{equation}
{\cal B}\left(B^0 \to \bar{D}^0 K^+\pi^-\right) = (9.0 \pm 0.6\,\text{(stat.)} \pm 0.7\,\text{(syst.)} \pm 0.9 ({\cal B}))\times10^{-5} \, ,\nonumber
\end{equation}
\begin{equation}
{\cal B}\left(B_s^0 \to \bar{D}^0 K^-\pi^+\right) = (1.00 \pm 0.04\,\text{(stat.)} \pm 0.10\,\text{(syst.)} \pm 0.10 ({\cal B}))\times10^{-3} \, ,\nonumber
\end{equation}
where the third uncertainty arises from the uncertainties on ${\cal B}(B^0 \to \bar{D}^0 \pi^+\pi^-)$.
This is the most precise measurement of ${\cal B}(B^0 \to \bar{D}^0 K^+\pi^-)$ to date and the first measurement of ${\cal B}(B_s^0 \to \bar{D}^0 K^-\pi^+)$.

\begin{figure}[htb]
\centering
\includegraphics[height=1.5in]{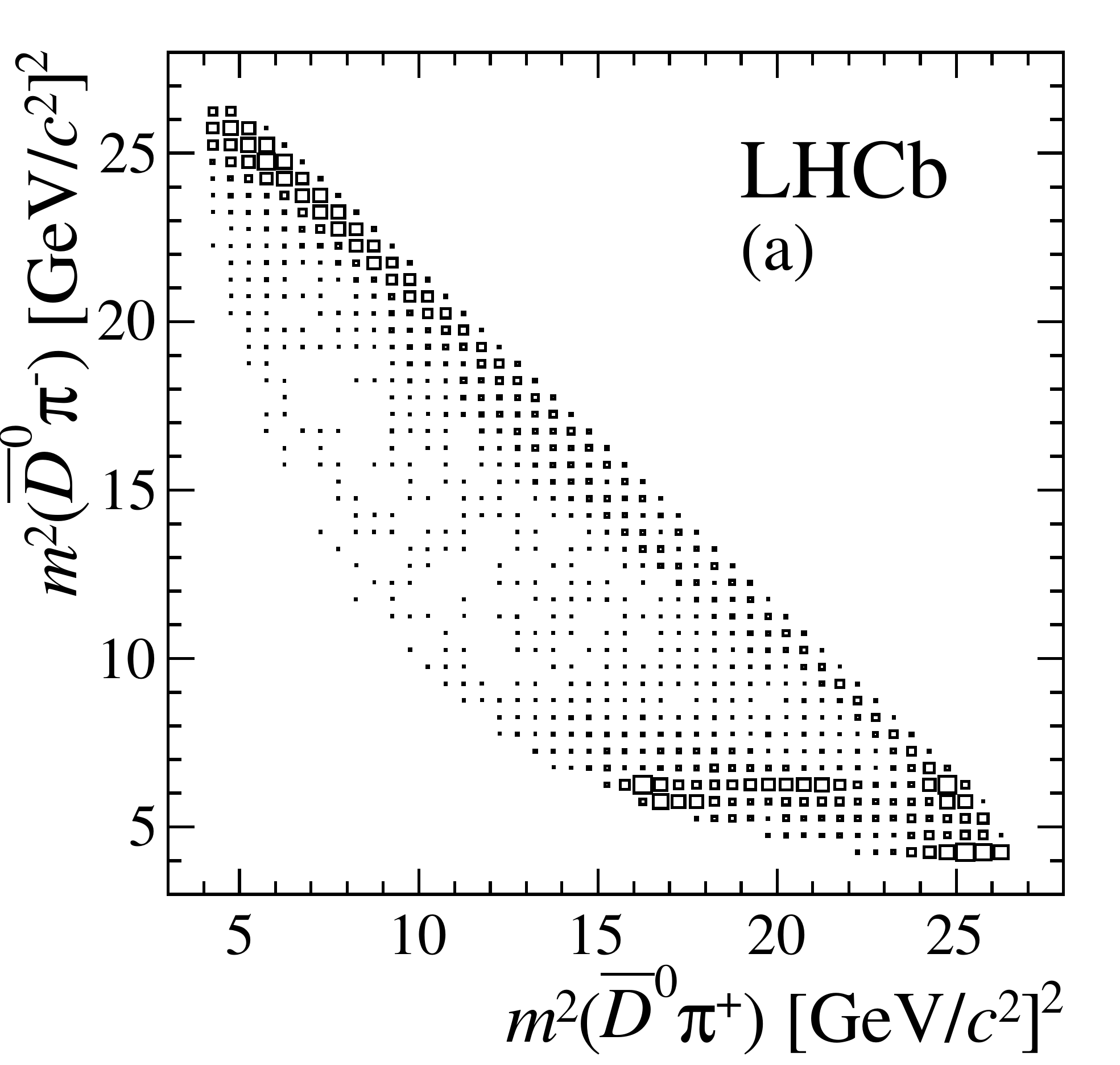}
\includegraphics[height=1.5in]{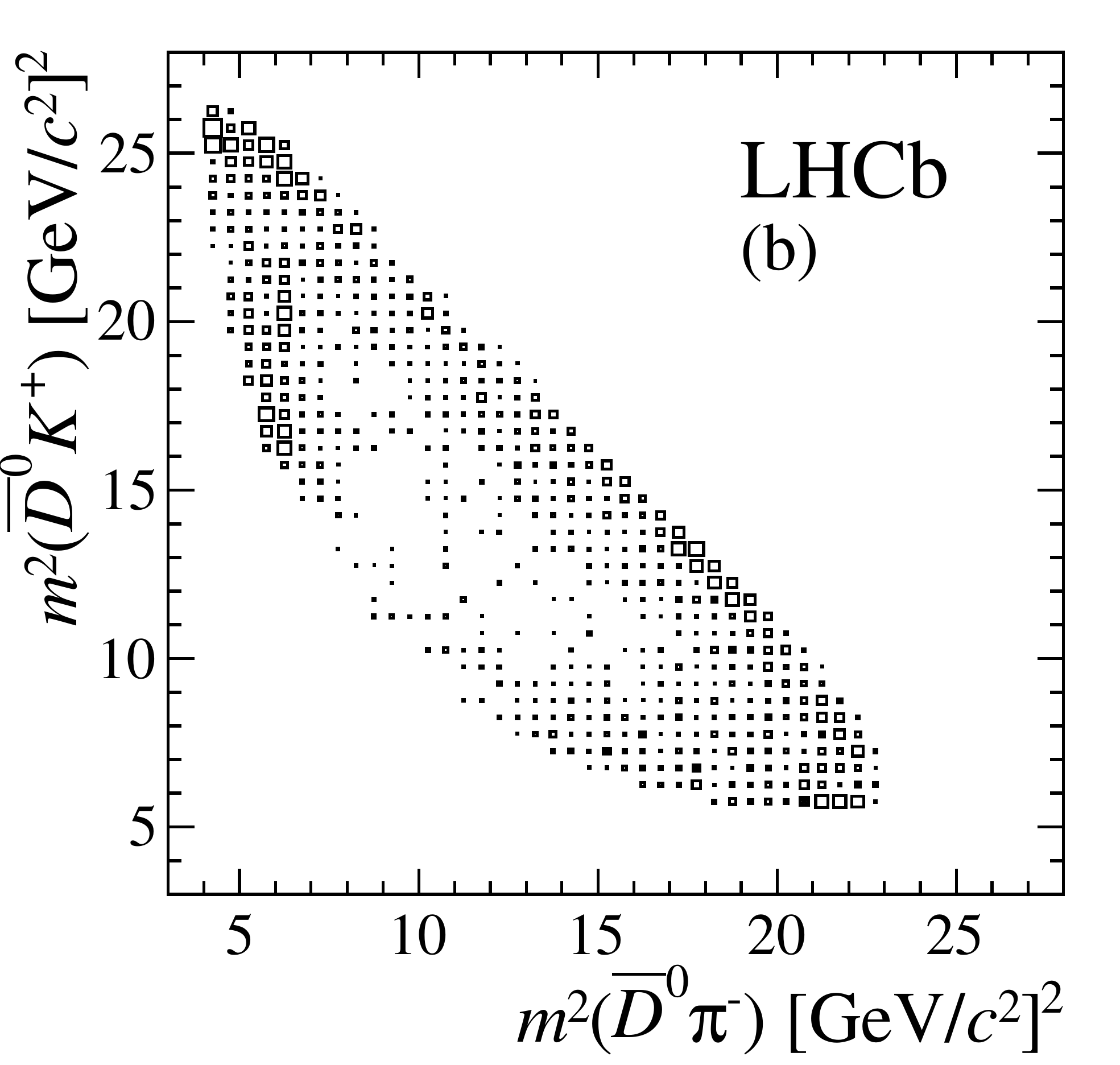}
\includegraphics[height=1.5in]{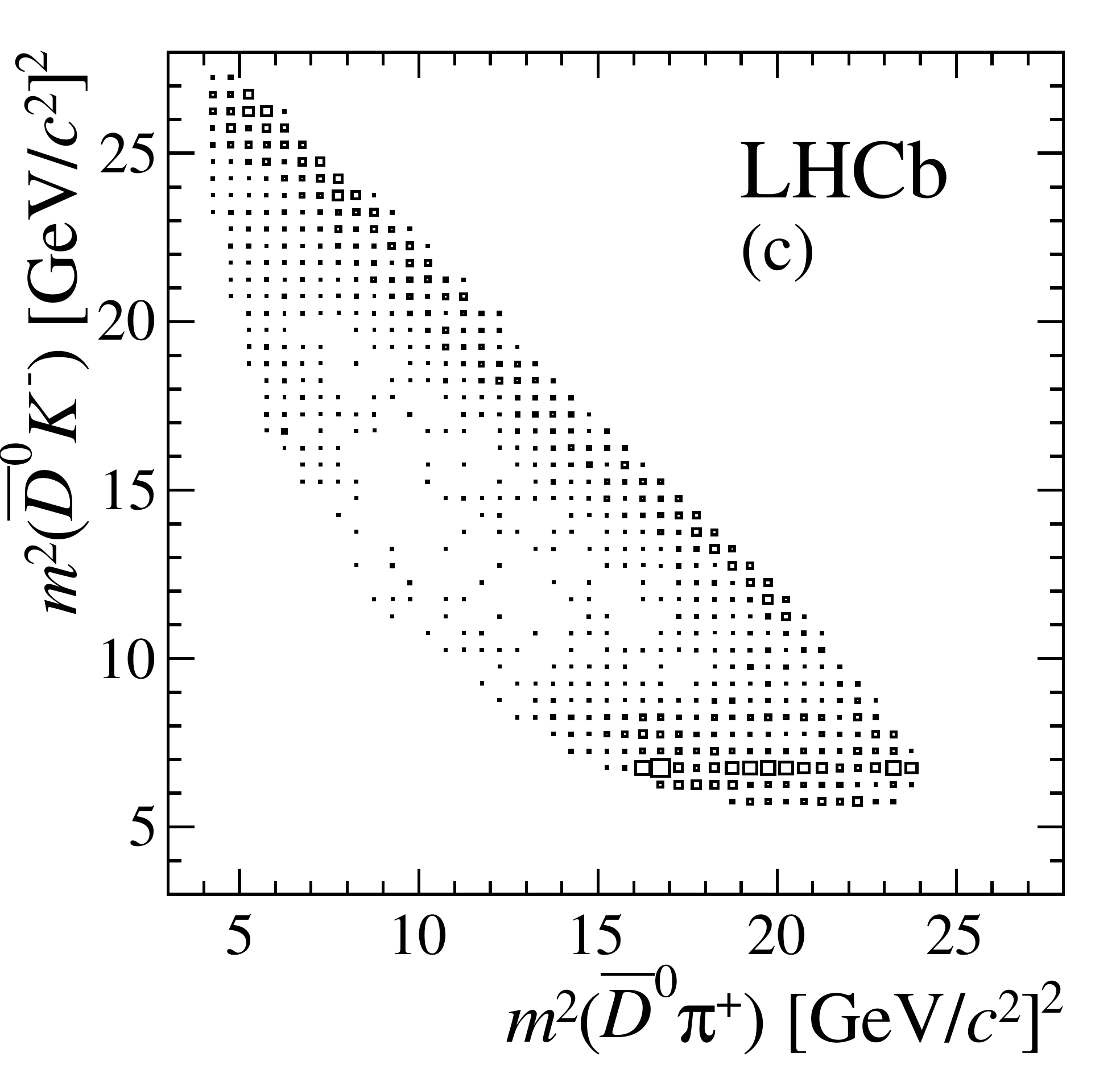}
\caption{Efficiency corrected Dalitz plot distributions for (a)~\mbox{$B^0\rightarrow\bar{D}^0 \pi^+ \pi^-$}, (b)~\mbox{$B^0\rightarrow\bar{D}^0 K^+ \pi^-$} and (c)~\mbox{$B_s^0\rightarrow\bar{D}^0 K^- \pi^+$} candidates obtained from the signal weights.}
\label{fig:dkpiDalitz}
\end{figure}

Although no quantitative analysis of the Dalitz plots has yet been attempted, the Dalitz plot distributions obtained 
(corrected for efficiency) are presented in Fig.~\ref{fig:dkpiDalitz}.

\section{Conclusions and prospects}

The $B^\pm \rightarrow DK^\pm$ decay mode offers an excellent opportunity to measure the CKM angle $\gamma$ from Standard Model processes. 
The combination in Section~\ref{sec:gamma:comb3fb-1} gives the most sensitive measurement of $\gamma$ from a single experiment so far, yielding a value of $(67 \pm 12)^\circ$.
This measurement is expected to improve further with the completion of a GLW/ADS analysis on the remaining $2\,\text{fb}^{-1}$ of LHCb data currently available.
In addition, other modes such as $B^0 \rightarrow D K^+ \pi^-$ offer great prospects for future $\gamma$ measurements. 

\Acknowledgments
This work is funded in part by the European Research Council under FP7 and by the United Kingdom's Science and Technology Facilities Council.

\end{document}